\documentclass[conference,compsoc]{IEEEtran}
\PassOptionsToPackage{hyphens}{url}

\usepackage{tabularx,booktabs}
\usepackage{graphicx}
\usepackage{amsmath}
\usepackage{array} 
\usepackage{url}
\usepackage{comment}

\usepackage{color,xcolor}
\usepackage{subcaption}
\usepackage{subfig}
\usepackage[]{hyperref}
\usepackage{array}
\usepackage{multirow}
\usepackage{multicol}
\usepackage{makecell}
\usepackage{cite}

\makeatletter
\def\sf@counterlist{}
\makeatother

\newcommand{\name}{InputSnatch}

\begin{document}

\title{\name: Stealing Input in LLM Services via Timing Side-Channel Attacks}

\thanks{*Xing Hu is the corresponding author (huxing@ict.ac.cn)}

\author{
    Xinyao Zheng\textsuperscript{1,2,3}, Husheng Han\textsuperscript{1,2,3}, Shangyi Shi\textsuperscript{1,2,3}, Qiyan Fang\textsuperscript{1,3,6}, Zidong Du\textsuperscript{1,5}, Xing Hu\textsuperscript{1,4}\textsuperscript{*},  Qi Guo\textsuperscript{1}\\
    \textsuperscript{1}\textit{SKLP, Institute of Computing Technology, Chinese Academy of Sciences}\\
    \textsuperscript{2}\textit{University of Chinese Academy of Sciences} \quad
    \textsuperscript{3}\textit{Cambricon Technologies} \\
    \textsuperscript{4}\textit{ZGC LAB} \quad
    \textsuperscript{5}\textit{Shanghai Innovation Center for Processor Technologies, SHIC} \\
    \textsuperscript{6}\textit{University of Science and Technology of China}\\
    
    \{zhengxinyao22s, hanhusheng20z, shishangyi22s\}@ict.ac.cn \\
    \{qiyanfang\}@mail.ustc.edu.cn\\
    \{duzidong, guoqi, huxing\}@ict.ac.cn \\}

\maketitle

\begin{abstract}
Large language models (LLMs) possess extensive knowledge and question-answering capabilities, having been widely deployed in privacy-sensitive domains like finance and medical consultation. During LLM inferences, cache-sharing methods are commonly employed to enhance efficiency by reusing cached states or responses for the same or similar inference requests. However, we identify that these cache mechanisms pose a risk of private input leakage, as the caching can result in observable variations in response times, making them a strong candidate for a timing-based attack hint.

In this study, we propose a novel timing-based side-channel attack to execute input theft in LLMs inference. The cache-based attack faces the challenge of constructing candidate inputs in a large search space to hit and steal cached user queries. 
To address these challenges, we propose two primary components. The input constructor employs machine learning techniques and LLM-based approaches for vocabulary correlation learning while implementing optimized search mechanisms for generalized input construction. The time analyzer implements statistical time fitting with outlier elimination to identify cache hit patterns, continuously providing feedback to refine the constructor's search strategy. We conduct experiments across two cache mechanisms and the results demonstrate that our approach consistently attains high attack success rates in various applications. Our work highlights the security vulnerabilities associated with performance optimizations, underscoring the necessity of prioritizing privacy and security alongside enhancements in LLM inference.

\end{abstract}


\IEEEpeerreviewmaketitle

\section{Introduction}

Since the release of ChatGPT~\cite{chatgpt}, large language models (LLMs) have garnered widespread attention due to their exceptional reasoning and natural language generation capabilities, leading to their extensive application in privacy-sensitive areas such as medical~\cite{llm_medical_survey}, finance~\cite{wu2023bloomberggpt}, law~\cite{cui2023chatlaw}, and office assistance~\cite{copilot}. These models demonstrate remarkable proficiency in understanding complex queries, generating contextually appropriate responses, and providing domain-specific insights across various fields. Users increasingly rely on these intelligent agents for privacy-critical scenarios like medical advice, financial planning, and legal consultation tasks. This growing dependence on LLM-based systems for processing sensitive personal information raises significant privacy and security concerns. 

LLMs' privacy and security are rapidly gaining attention as these models become increasingly integrated into commercial applications. Building upon traditional deep learning privacy concerns, the enhanced memorization capabilities of LLMs exacerbate existing privacy risks such as membership inference attacks (MIAs)\cite{song2019auditing, mattern2023membership, meeus2024did,carlini2022quantifying} and personally identifiable information (PII) leakage\cite{lukas2023analyzing, kim2024propile, staab2023beyond}. These risks scale proportionally with model size, as larger models demonstrate stronger memorization of training data~\cite{carlini2022quantifying}.

Beyond these traditional concerns, LLMs introduce a novel privacy challenge:  prompt theft attacks, which threaten intellectual property rights and personal privacy. These prompt theft attacks have emerged in various forms, including extracting system prompts by adversarial prompting~\cite{zhang2024effective, yu2023assessing, liang2024my, perez2022ignore}, inverting prompts from embedding vectors~\cite{song2020information,morris2023text, chen2024text,li2023sentence,chen2024text} or model responses~\cite{shen2024prompt, mahajan2024prompting, sha2024prompt, yang2024prsa}, and recovering input prompts by exploiting next-token probability distributions~\cite{morris2023language} or token-length sequences~\cite{Keylogging_prompt}. These attacks pose significant risks to the growing prompt marketplace and user privacy in downstream LLM applications, particularly when prompts contain sensitive information or proprietary instructions.

Although previous works have demonstrated the prompt theft attacks in LLM-based systems, they have several limitations:  1) Inability to recover exact privacy:  current approaches rely on either response feature analysis~\cite{shen2024prompt, yang2024prsa, prompt_inversion} or embedding vectors~\cite{song2020information, chen2024text}, can only achieve partial semantic recovery. This is attributed to the complexity of establishing precise inverse mappings in high-dimensional spaces, resulting in the loss of crucial information. 2) Limited attack scenarios:  existing attacks are designed for specific applications based on task-specific methodologies~\cite{prompt_inversion,shen2024prompt, yang2024prsa}. While effective in their targeted domains, these specialized approaches can not transfer to other LLM-based applications, leading to limited attack scenarios. 3) Impractical assumptions:  current attacks often rely on unrealistic assumptions, such as white-box access to models~\cite{song2020information} or unlimited querying capabilities to obtain next-token probability distributions across the entire vocabulary~\cite{morris2023language}.

\begin{figure}[!htbp]
    \centering
    \includegraphics[width=0.95\linewidth]{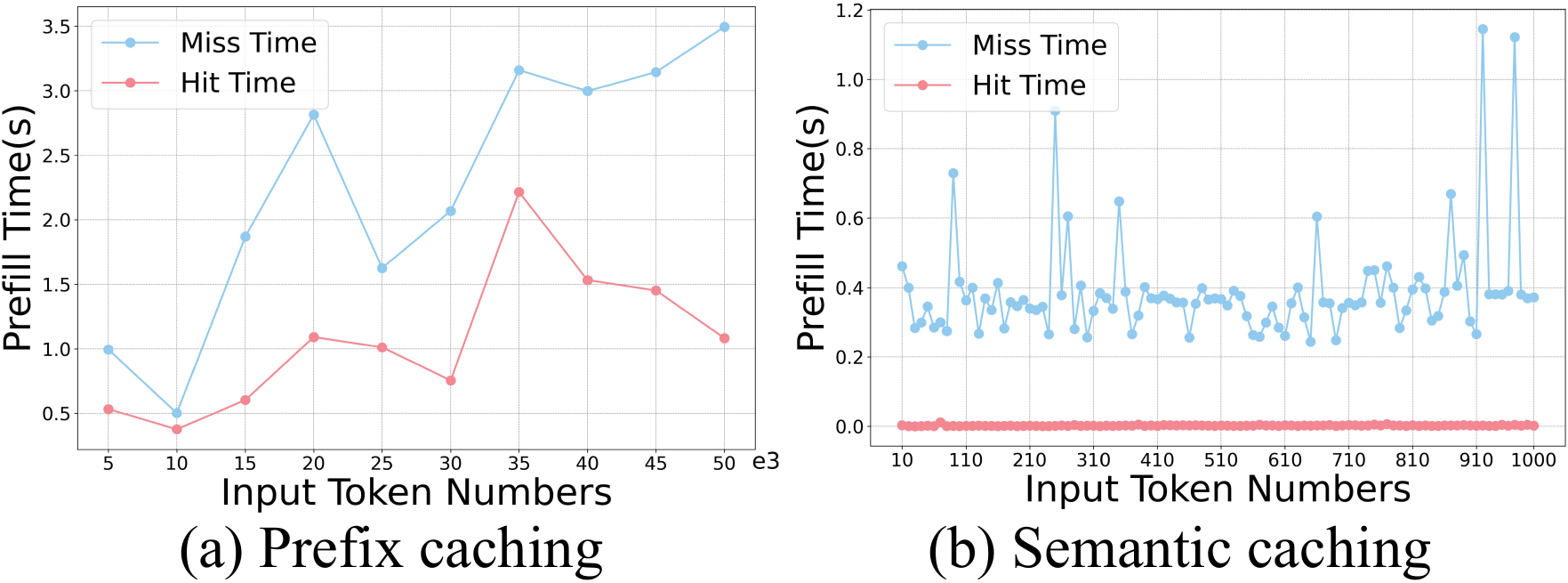}
    \caption{The prefill time difference between cache hits and misses for varying input lengths with OpenAI API calls GPT-4o-mini LLM. (a) Prefix caching implemented by OpenAI. (b) Semantic caching with GPTCache.}
    \label{fig:  latency_comp}
\end{figure}

In this work, we present \name, which addresses the limitations of existing attacks by exploiting timing-based side channels arising from the commonly used cache-sharing optimization (prefix caching and semantic caching). Prefix caching~\cite{kwon2023efficient, kwon2023efficient, ye2024chunkattention, juravsky2024hydragen} enables the reuse of cached attention states in two different requests with the same prefix prompts, having been widely deployed in the industry environment. Semantic caching~\cite{bang2023gptcache, mohandoss2024context, li2024scalm,gill2024privacy} allows requests with identical content or similar semantics to share cached responses, especially in LLM applications equipped with Retrieval-Augmented Generation(RAG).
\textit{Our primary observation is that different requests on the same computing nodes share the same cache, which results in an observable time reduction when meeting sharing criteria.}
Figure~\ref{fig:  latency_comp} illustrates a pronounced time difference between hits and misses, despite fluctuations caused by schedule issues and network variability. The observable time differences provide attack hints for attackers to obtain cache states and infer secret inputs of other users, potentially disclosing personal privacy information, trade secrets, or other enterprise data.
Noted that, our proposed cache-based attack vector is powerful because (1) it enables exact input reconstruction due to the strict matching requirements in prefix caching, (2) it generalizes well for its wide implementation in diverse mainstream LLM inference backends~\cite{Huggingfacetextgenerationinference, NVIDIATensorRTLLM, Huggingfacetextgenerationinference} and API providers~\cite{OpenAIPrefixcaching, anthropicPromptCaching, googleContextCaching, microsoftTutorialAzure, amazonsemanticcaching}, (3) it is practical since no privileged access beyond standard API capabilities is required. 



The cache-based attack requires input construction for the desired cache hit. For this purpose, our attack framework comprises two components:  a \textit{Input Constructor}, which generates inputs attempting to hit cached content, and the \textit{Time Analyzer}, which determines whether a match has occurred based on the measured time. However, fully recovering prompts present several significant challenges. First, the complexity of input construction escalates exponentially as the vocabulary size reaches 1000,000 tokens~\cite{tao2024scaling} and the context window scales to 128K ~\cite{gpt128k}, substantially increasing the search space. Second, the observed times are influenced by noise coming from network latency and memory scheduling delays, complicating accurately determining the cache state. Moreover, practical deployment constraints such as available memory, rate limit, and cache's Time To Live (TTL), pose additional challenges to our attack.


To address these challenges, we introduce a comprehensive timing analysis framework. First, the framework establishes temporal patterns of target services by strategic sampling method and mitigates noise interference with proposed point processing algorithms. Second, to mitigate the large search space issue, our framework leverages advanced machine learning techniques to extract contextual information and semantic relationships from open-source datasets, thereby enhancing construction efficiency. Besides, we incorporate a multi-stage candidate evaluation process to prioritize candidates exhibiting higher cache hit probabilities. In the process, the constructed texts undergo adaptive filtering mechanisms and probabilistic ranking algorithms to optimize the selection of high-potential candidates while maintaining computational efficiency.

Experimental evaluations across diverse deployment scenarios demonstrate the effectiveness of our attack framework. For requests of specified lengths, our timing analyzer achieves 87.13\% accuracy in determining cache hit prefix lengths. In medical question-answering systems with static prompt engineering, our attacks achieve a 62\% success rate in extracting exact disease inputs and 13.5\% for precise symptom descriptions. For legal consultation services implementing RAG, the semantic extraction success rates range from 43\% to 100\%. These results highlight significant privacy vulnerabilities across deployment contexts, emphasizing the critical balance between performance optimization and security in LLM services.

\noindent\textbf{Contributions.} The main contributions of this work are: 
\begin{itemize}
    \item First systematic investigation of time-based side channels in LLM inference caused by cache-sharing optimization, analyzing privacy leakage risks of two cache mechanisms and examining their performance-privacy trade-offs.
    \item Implementation of a comprehensive attack methodology combining diverse input construction strategies (ML models, LLM-based analysis, and optimized search) with robust timing analysis (statistical fitting and anomaly detection), demonstrating effective input reconstruction across various deployment scenarios.
    \item Development of a robust attack framework that achieves 62\% success rate in exact partial input recovery and 12.5\% in exact complete input extraction, while demonstrating 79.5\% effectiveness in semantic-level content reconstruction under real-world deployment constraints. 
    
\end{itemize}

\section{Background}
\subsection{Transformer-Based LLM Inference}

Transformer-based large language models (LLMs) such as GPT-3~\cite{brown2020language} and PaLM~\cite{chowdhery2022palm} have revolutionized natural language processing by leveraging the transformer architecture~\cite{vaswani2017attention}. A core component of these models is the self-attention mechanism, which enables the capture of long-range dependencies and complex contextual relationships within input sequences. 

The self-attention mechanism mainly contains the projection and attention operations. Given a token sequence $\mathbf{X} = [\mathbf{x}_1, \dots, \mathbf{x}_n]$, each token $\mathbf{x}_i$ is transformed into query ($\mathbf{q}_i$), key ($\mathbf{k}_i$), and value ($\mathbf{v}_i$) vectors through learned linear projections as $\mathbf{q}_i = \mathbf{W}_q \mathbf{x}_i, \quad \mathbf{k}_i = \mathbf{W}_k \mathbf{x}_i, \quad \mathbf{v}_i = \mathbf{W}_v \mathbf{x}_i$, where $\mathbf{W}_q$, $\mathbf{W}_k$, and $\mathbf{W}_v$ are trainable weight matrices. Then, the attention operations with Q, K, and V are processed as shown in Euqation~\ref{eq:  attention}. After this, a linear layer performs re-projection to generate the input for the next layer. The computational complexity of self-attention increases quadratically with the length of the tokens.

\begin{equation}
\label{eq:  attention}
Attention(\mathbf{Q}, \mathbf{K}, \mathbf{V}) = \text{softmax}\left(\frac{\mathbf{Q} \mathbf{K}^\top}{\sqrt{d_k}}\right)\mathbf{V}
\end{equation}

LLMs generate text autoregressively, producing one token at a time conditioned on the initial prompt and the preceding generated tokens.
The generation process typically involves two phases:   \textit{prefill} and \textit{decoding}. 
The prefill phase computes input prompts intermediate states (keys and values) for input prompts and generates the first next token.
The decoding phase generates output tokens autoregressively one by one until a stopping token is met.
Due to data dependencies, the autoregressive generation cannot be parallelized, resulting in the underutilization of GPU resources and memory constraints, contributing significantly to the latency of individual requests. It's imperative to enhance metrics like Time To First Token (TTFT) and Time Per Output Token (TPOT), as latency can be calculated as $Latency = TTFT + (TPOT * N)$, where $N$ represents the number of generated tokens.

\subsection{Cache Optimization in LLM Inference}

\begin{figure}[!htbp]
    \centering    
    \includegraphics[width=0.95\linewidth]{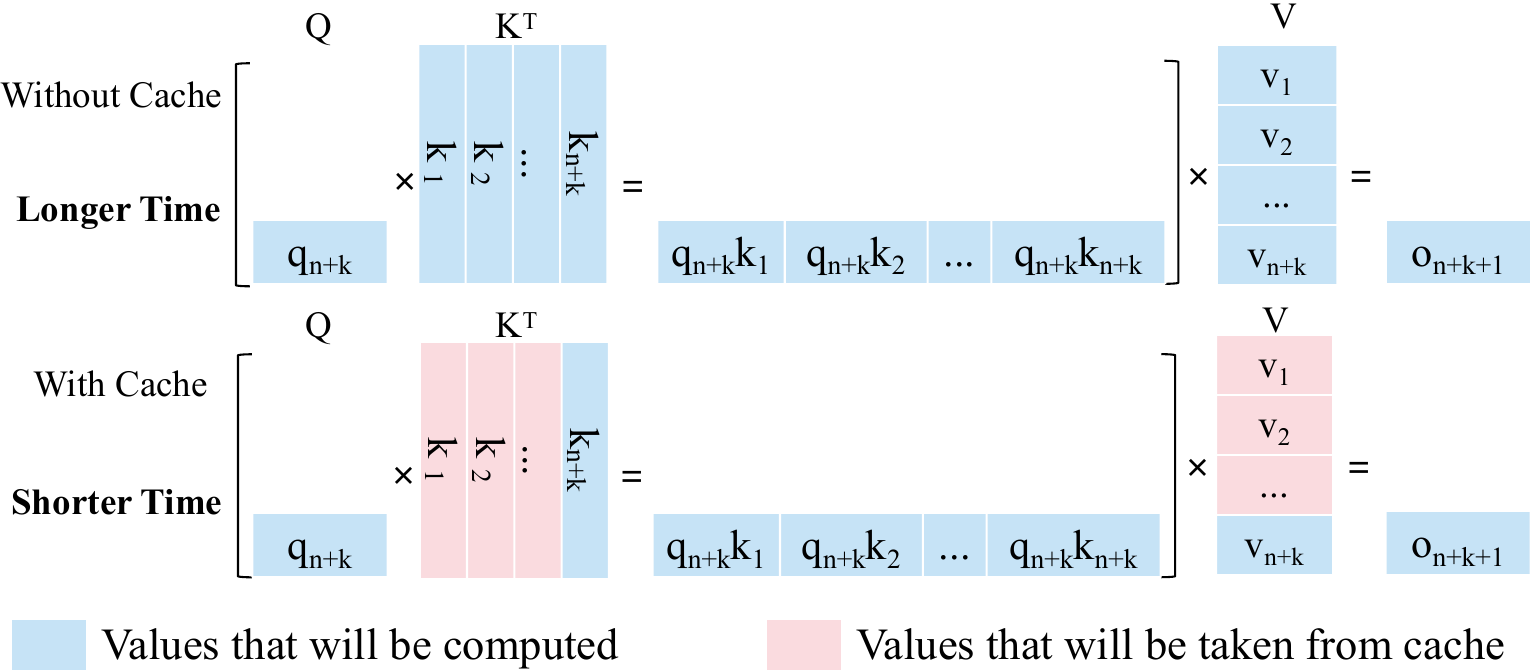}
    \caption{Comparison of self-attention computation mechanisms. The traditional approach (upper) performs full recomputation for each token, while the KV cache (lower) reuses stored key-value vectors to accelerate inference. The KV Cache reduces the computational complexity per decoding step from $O(n^2)$ to $O(n)$. }
    \label{fig:   kvcache}
\end{figure}

LLM inference services deployed on the cloud resources require managing a high volume of real-time requests while ensuring high throughput and low latency. The concurrent requests are scheduled across different computing nodes and computed in optimized batching~\cite{daniel2023continuous,nvidiaRaggedBatching} to increase throughput. Modern LLMs utilize the Key-Value (KV) Cache mechanism~\cite{pope2023efficiently} to optimize TPOT by storing each token's key and value vectors after their initial computation. As presented in Figure~\ref{fig:   kvcache}, when generating token vector $o_{n+k+1}$, the attention states of previous tokens can be cached to avoid recomputation. However, as the sequence length and batch size increase, the KV cache consumes more GPU memory. Many inference optimizations technologies have been proposed to optimize KV Cache, including sparsity~\cite{zhang2024h2o,dong2024get,zandieh2024subgen}, quantization~\cite{dettmers2022gpt3,frantar2022gptq,yue2024wkvquant,lin2024awq,shao2023omniquant}, windowing~\cite{han2023lm,xiao2023efficient}, and sharing~\cite{kwon2023efficient,juravsky2024hydragen,zheng2023efficiently,ye2024chunkattention}. Leading LLM API providers extensively implement cache-sharing mechanisms (detailed in Table~\ref{tab:  api_vendors}), which can be classified into prefix and semantic caching approaches, to optimize TTFT and memory utilization through cross-request content reuse.

\begin{table}
    \centering
    \caption{Cache Mechanisms Comparison of Different LLM API Vendors}
    \label{tab:  api_vendors}
    \resizebox*{0.98\linewidth}{0.17\textheight}{
    \begin{tabular}{lcll} 
        \toprule
        Vendor          & Stream & Caching Mechanisms          & Cache Lifetime   \\ \midrule
        OpenAI~\cite{OpenAIPrefixcaching}  & Y      & Prefix Caching    & 5-10 minutes     \\
        DeepSeek~\cite{DeepSeekapi}        & Y      & Prefix Caching    & Hours to days    \\
        Anthropic Claude~\cite{anthropicPromptCaching} & Y      & Prefix Caching    & 5 minutes        \\
        Google Gemini~\cite{googleContextCaching}   & Y      & Prefix Caching    & Default 1 hour    \\
        MoonShot Kimi~\cite{MoonShotAPI}       & Y      & Prefix Caching    & Customization    \\ 
        Portkey~\cite{portkeysemantic}         & Y      & Semantic Caching  & Default 7 days   \\
        Google Cloud~\cite{mediumImplementingSemantic}    & Y      & Semantic Caching  & Uncertain        \\
        Microsoft Azure~\cite{microsoftTutorialAzure} & Y      & Semantic Caching  & Uncertain        \\
        UnKey~\cite{unkeySemanticCaching}           & Y      & Semantic Caching  & Uncertain        \\
        Amazon~\cite{amazonsemanticcaching}          & Y      & Semantic Caching  & Uncertain        \\
\bottomrule
    \end{tabular}
    }
\end{table}

\noindent\textbf{Prefix Caching.} Prefix caching~\cite{zheng2023efficiently, kwon2023efficient, ye2024chunkattention, juravsky2024hydragen} reuses the KV cache with identical prefixes across different requests or within multiple sequences generated from a single request. For instance, vLLM~\cite{kwon2023efficient} introduces non-contiguous memory allocation and efficient prefix sharing across different sequences. It is important to note that only consistent content from the beginning qualifies as a prefix; the same segments starting mid-sequence cannot be matched. Prefix caching has been an industry norm integrated into mainstream inference frameworks such as HuggingFace TGI~\cite{Huggingfacetextgenerationinference}, NVIDIA TensorRT-LLM~\cite{NVIDIATensorRTLLM}, and LMDeploy TurboMind~\cite{lmdeploy}. Other notable implementations include SGLang~\cite{zheng2023efficiently}, which employs a radix tree to reuse the KV cache in various scenarios, ChunkAttention~\cite{ye2024chunkattention}, which introduces a prefix-aware KV cache partitioning mechanism, and Hydragen~\cite{juravsky2024hydragen}, which efficiently manages shared prefixes. 

\noindent\textbf{Semantic Caching.} Semantic caching mechanism~\cite{bang2023gptcache, mohandoss2024context, li2024scalm, gill2024privacy} enables consistent outputs for identical or semantically similar queries. This approach optimizes server resource utilization, minimizes data retrieval latency, reduces API costs, and enhances system scalability. By leveraging embedding similarity metrics~\cite{zhu2023optimal, zhuefficient}, responses can be reused for semantically equivalent queries. A prominent open-source implementation, GPTCache~\cite{bang2023gptcache}, demonstrates this concept by caching request-response pairs. Upon cache hits, the system directly retrieves stored responses, reducing redundant LLM invocations and response latency. Recent research efforts~\cite{mohandoss2024context,li2024scalm} have focused on enhancing semantic caching optimization strategies to improve cache hit rates and minimize operational costs.

However, these caching mechanisms are potential attack targets, as different users' requests can share the same caching infrastructure. Our focus is on streaming responses supported by the APIs, where generated tokens are sent back sequentially, thus enabling users to receive real-time feedback and measure per-token generation times precisely. We have designed attack strategies targeting two scenarios, aiming to extract information about input prompts by measuring timing side channels associated with the cache mechanisms. Specifically, we focus on two application scenarios:   static prompt engineering using prefix caching and RAG  employing semantic caching. By timing side-channel attacks, we can infer partial or complete information about user inputs, posing a security threat as unauthorized access to sensitive input information.

\subsection{Targeted Attack Scenarios}

Many applications leverage LLMs' advanced reasoning and natural processing capabilities by customizing models for specific tasks. They often combine technologies such as prompt engineering or Retrieval-Augmented Generation (RAG) at a high level, offering advanced functionality and exceptional adaptability across various industries, including healthcare~\cite{singhal2023towards,tu2024towards,singhal2023large,yang2022gatortron,sallam2023chatgpt}, legal consulting~\cite{cui2023chatlaw,shi2024legal,wu2024knowledge,sun2024lawluo
}, e-commerce~\cite{fang2024llm,yang2023mixpave,zou2024implicitave,guo2023comave
}, and education~\cite{wen2024ai,bui2024cross,milano2023large,lieb2024student}.

\noindent\textbf{Prompt Engineering.} Prompt engineering has become indispensable for extending the model's capabilities through task-specific instructions, allowing integration into downstream tasks ~\cite{sahoo2024systematic}. These techniques include few-shot prompting~\cite{brown2020language,gao2020making,gu2022ppt}, where examples guide the model's responses for new tasks without extensive training; chain-of-thought prompting~\cite{wei2022chain,zhang2022automatic,zhao2023enhancing,li2023structured}, which encourages step-by-step reasoning. Recent studies have shown that carefully crafted prompts can significantly improve model performance across various tasks, from question answering~\cite{wang2023layout,tan2023make,shao2023prompting,singhal2023towards} to code generation~\cite{doderlein2022piloting,chen2024evoprompting}. For example, ChatGPT o1~\cite{chatgpto1} leverages reinforcement learning to refine its CoT strategies, allowing it to identify and correct errors effectively. 

Our attack strategy focuses on applications that use static prompt engineering, like GPT store~\cite{openaiExploreGPTs}, where users can design system prompts for specific tasks to define their apps. User input is usually embedded or concatenated into the system prompt, which can be reused with prefix caching. It's possible to recover partial or complete input accurately due to KVCache segmented management and precise prefix matching deployed by mainstream inference frameworks (as detailed in section 4.2).

\begin{figure}[h]
    \centering    
    \includegraphics[width=0.95\linewidth]{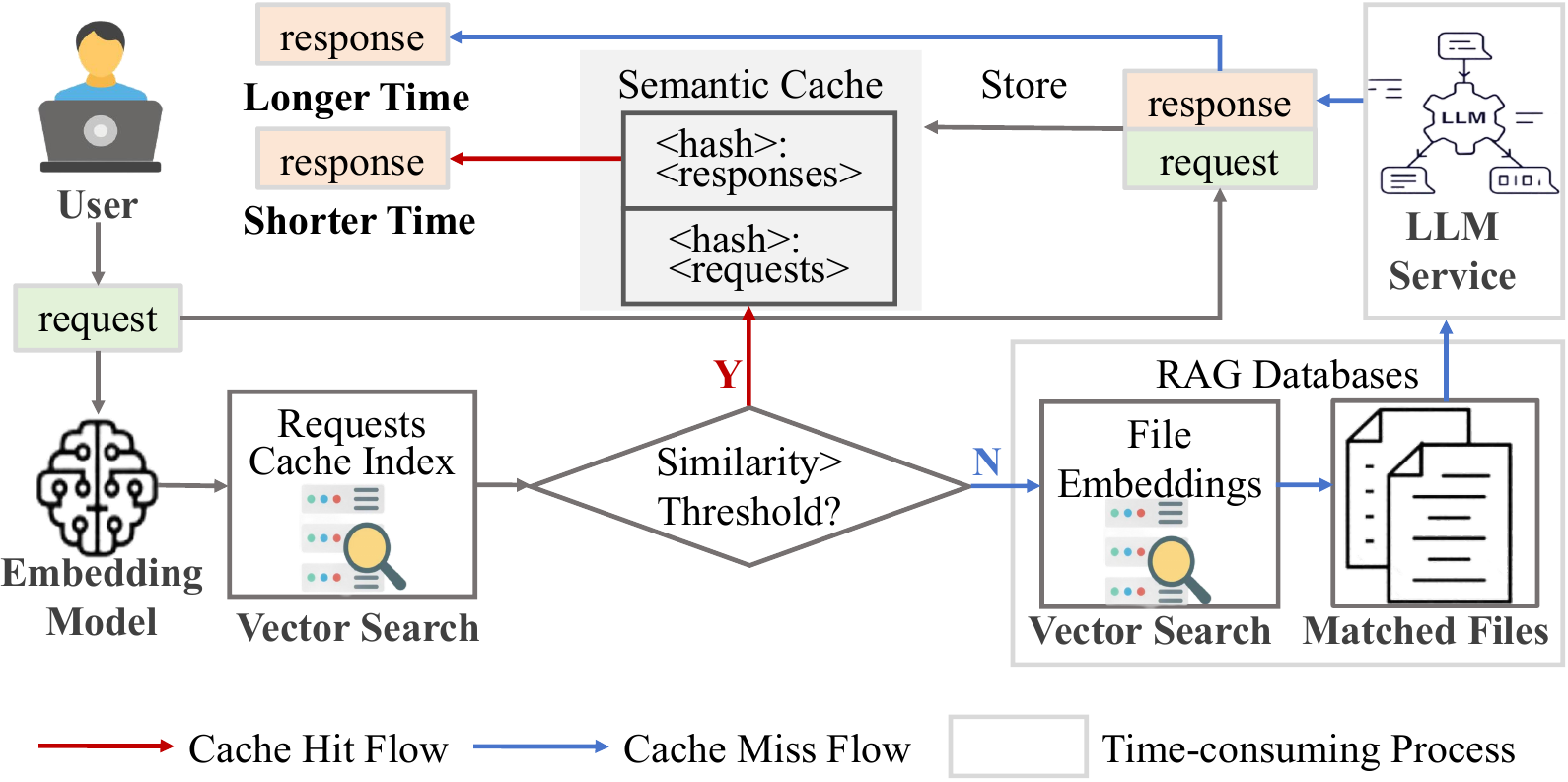}
    \caption{Overview of the RAG-assisted LLM system with the semantic caching mechanism. User queries are first matched against cached requests based on semantic similarity. Responses are retrieved directly from the cache if the similarity score exceeds the threshold; otherwise, the system proceeds with vector database retrieval and LLM inference. }
    \label{fig:   RAG}
\end{figure}

\noindent\textbf{Retrieval-Augmented Generation (RAG).} RAG\cite{lewis2020retrieval} enhances LLMs by retrieving relevant external knowledge to address the inherent limitations of LLMs, including knowledge constraints and hallucination issues. This approach has demonstrated significant value across various domains, including improving accuracy in medical diagnostics\cite{xia2024rule} and legal analysis~\cite{kalra2024hypa} and integrating external information in open-domain questions~\cite{wang2023learning,nguyen2024sfr}. However, each interaction requiring data retrieval can be computationally intensive and time-consuming, particularly with large datasets. 

Developers exploit semantic caching to solve latency and cost challenges. As illustrated in Figure~\ref{fig:   RAG}, incoming user requests are translated into vectors by embedding models and compared against request embedding vectors through semantic similarity computation. If the similarity score exceeds the predefined threshold, the system directly retrieves the corresponding response from the semantic cache, bypassing the time-consuming retrieval and inference stages. When no semantically similar cached queries are found, the system follows the traditional pipeline:   retrieve relevant passages from the vector database, which are combined with the original query vectors and fed into LLMs for response generation. 

This semantic caching mechanism significantly reduces response latency for frequently asked questions. The distinct temporal patterns between cache hits and misses create an observable timing differential, as caching hits mitigate costly retrieval and generation operations latency. This discrepancy in response times potentially reveals sensitive information about the existence and nature of previously cached requests.

\section{Attack Hints and Threat Model}

\subsection{Attack Hints}

LLM inference systems leverage caching mechanisms to optimize computation, while the cache-sharing strategy inadvertently introduces vulnerabilities for time-based side-channel attacks. As demonstrated in Figure~\ref{fig: latency_comp}, despite significant timing variations due to network latency and scheduling delays, there exists evident time differences in real-world API deployments. In this section, we analyze the timing characteristics caused by two cache mechanisms at different phases (prefill and decode phase) to determine the specific attack vector and attack timing. We adopted two prominent open-source implementations:  vLLM~\cite{kwon2023efficient} and GPTCache~\cite{bang2023gptcache}  as our experiment platforms, which were chosen based on their widespread adoption, active development, and robust technical implementations.

\begin{figure}[!htbp]
    \centering
    \includegraphics[width=0.95\linewidth]{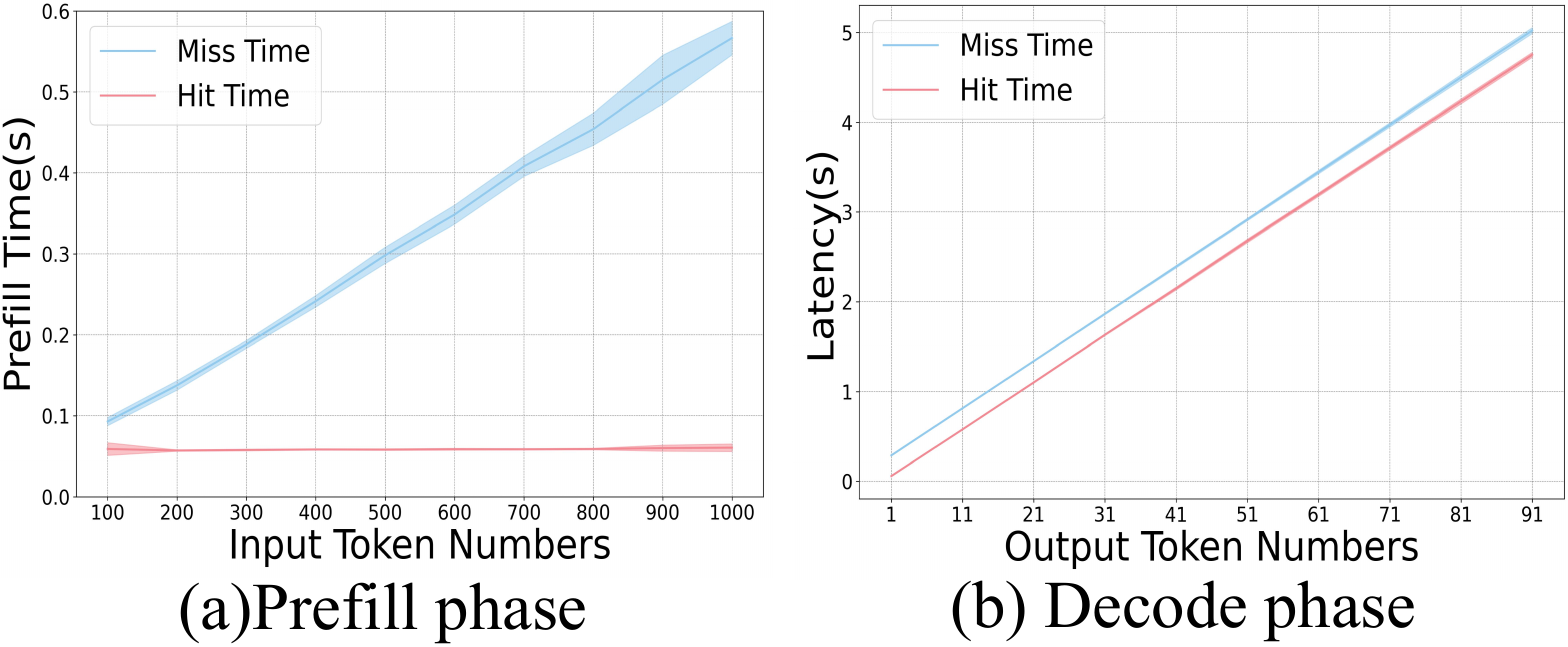}
    \caption{Time difference between hits and misses for prefix caching:  100 experiments in vLLM by a local API deployment using the LLaMa-2 70B model. (a)Time for input with varying lengths taken to generate one token. (b)Time for input with the same length to generate different numbers of tokens.}
    \label{fig: vllm_exp}
\end{figure}

Our empirical analysis revealed distinct temporal disparities between prefix cache hits and misses in the prefill and decode phases. Based on this observation, we selected the prefill time as an effective attack vector as it enables early termination of the inference process after the first token generation, significantly reducing the overall attack duration. As illustrated in Figure~\ref{fig:  vllm_exp}, our experiments reveal distinct timing patterns between cache hits and misses with no overlapping or ambiguous cases observed. Figure~\ref{fig:  vllm_exp}(a) shows that the timing difference in the prefill phase positively correlates with the length of cached requests. Figure~\ref{fig: vllm_exp}(b) demonstrates that the timing difference in the decode phase primarily stems from the time saved by reusing input token computations.


\begin{figure}[!htbp]
    \centering
    \includegraphics[width=0.95\linewidth]{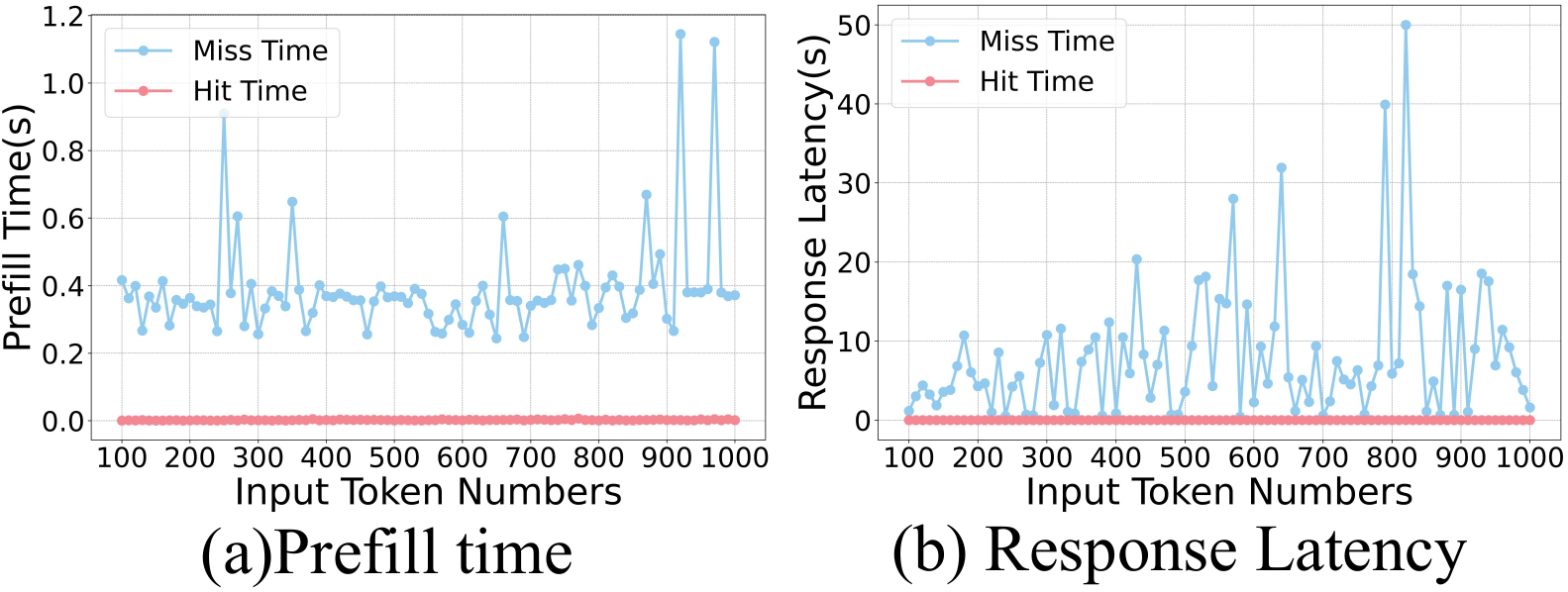}
    \caption{Time difference between hits and misses for semantic caching:  conducted in GPTCache by invoking API to access GPT-4o-mini. (a)Time for different inputs with varying lengths to generate one token. (b)Time for input with varying lengths to generate complete responses.}
    \label{fig:  gptcache_exp}
\end{figure}

Similarly, we identify distinguishable temporal patterns in semantic caching operations and select prefill time as the attack vector. When cache hits occur, the responses are retrieved from the cache, eliminating the need for token-by-token generation. This results in negligible prefill time and overall latency, as demonstrated in Figure~\ref{fig:  gptcache_exp}.  For cache misses, the response latency varies dramatically for different generated tokens, while the prefill time fluctuates slightly due to network latency.


\begin{figure*}[!htbp]
    \centering
    \includegraphics[width=0.95\linewidth]{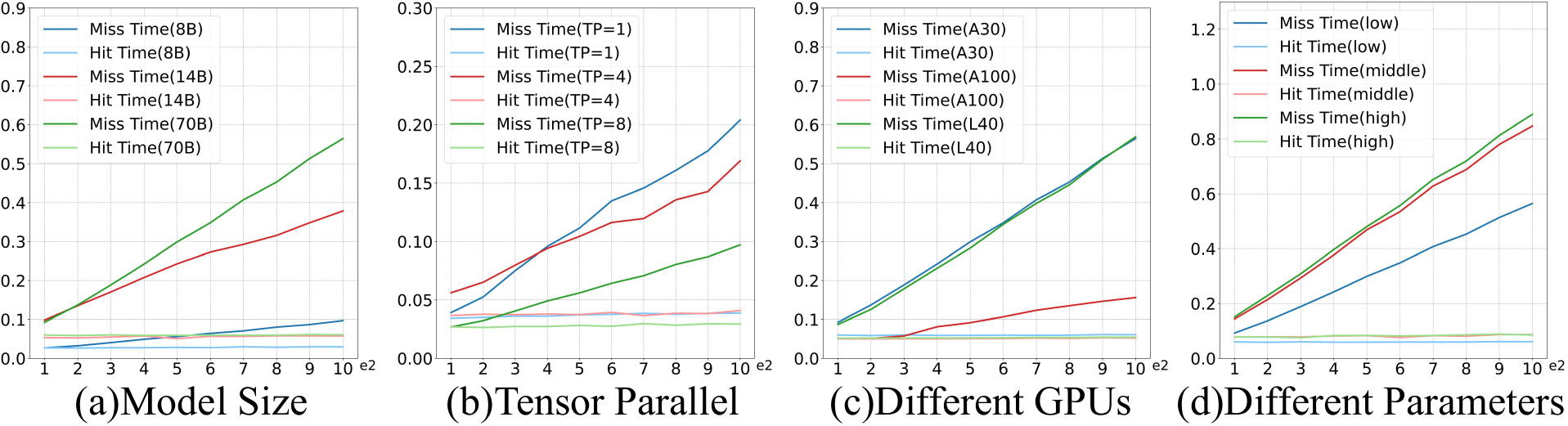}
    \caption{Prefill time characteristics under different configurations. The horizontal axis represents the number of input tokens, and the vertical axis represents the prefill time(s). (a) Three models of different parameter scales (LLaMa-3 8B~\cite{dubey2024llama}, Qwen 14B~\cite{bai2023qwen}, and LLaMa-2 70B~\cite{touvron2023llama}) deployed on eight A30 GPUs. (b) LLaMa-3 8B under varying tensor parallelism sizes (1, 4, and 8) across different numbers of A30 GPUs. (c) LLaMa-2 70B deployed across eight different GPU architectures (A30, A100, and L40). (d) LLaMa-2 70B deployed across eight A30 with different sampling diversity:  low (temperature=0, top\_k=1, top\_p=1), medium (temperature=0.2, top\_k=2, top\_p=0.3), and high (temperature=1.0, top\_k=100, top\_p=1.0).}
    \label{fig:  observation}
\end{figure*}

Given the clear time difference in both prefix and semantic caching mechanisms, we selected prefill time as our attack vector, allowing us to minimize the total attack duration. We conducted further experiments across various configurations as Figure~\ref{fig:  observation}, with each data point representing the average of 30 runs on vLLM. While our results indicate that TTFT correlates with model size, hardware specifications, tensor parallelism scale, and sampling parameters, a significant timing differential between hit and miss persists across all experimental configurations. This insight enables us to construct a robust time analyzer that can effectively characterize and differentiate caching behaviors, regardless of the underlying model architecture, hardware configurations, or sampling parameters in inference deployments.

\subsection{Threat Model}

\begin{figure}[!htbp]
    \centering    
    \includegraphics[width=0.95\linewidth]{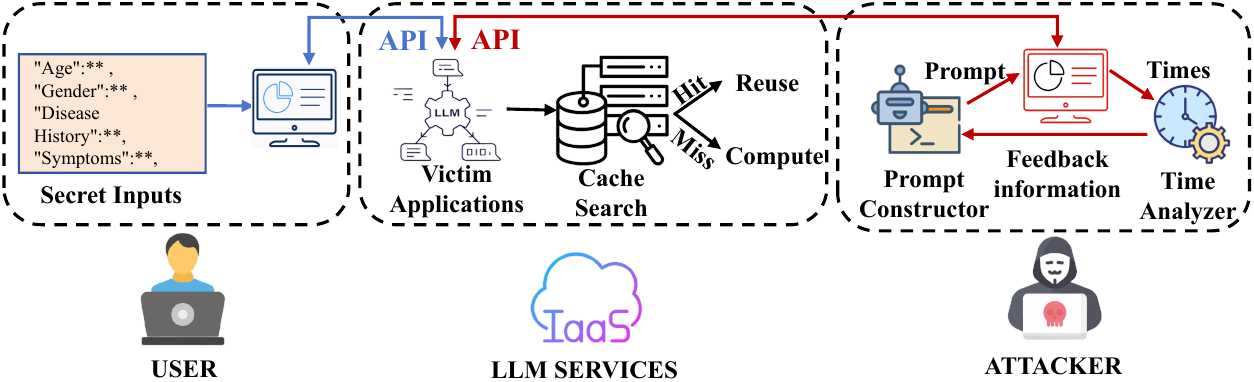}
    \caption{Overview of our attack scenario. Users interact with the LLMs cloud service through the interaction interface, and different user requests are routed to share the cache through encrypted API channels.}
    \label{fig:  attack_model}
\end{figure}

\noindent\textbf{Attack Scenario. }Our attack scenario is presented as Figure~\ref{fig:  attack_model}, the application was customized by developers to invoke the LLMs inference service provided by the cloud service vendor through the API. Ordinary users interact with the inference service through a web browser or other application interface, sending their privacy input and receiving generated responses. The communication channels can be encrypted to ensure data security. 

We consider the cloud infrastructure often implements shared caching mechanisms in each computing node, where requests from multiple users share cached states or responses. When cache hits occur, the computing nodes can bypass computation and directly use existing cached values to improve throughput and reduce latency. The experimental analysis demonstrates that systems with long shared prefixes achieve a 70\%-90\% reduction in memory utilization~\cite{ye2024chunkattention} and semantic caching can effectively handle 20-30\% of user queries~\cite{microsoftOptimizeAzure} in production deployments.

\noindent\textbf{Attacker’s Objective. }The primary objective of the attacker is to construct input to match inputs from normal users on the same computing node. With measured response latency, the attacker can determine the occurrence of cache hits, indicating the successful match of another user's input. For context retention and reuse maximization, requests from the same user are typically routed to the same computing node, which enables multiple attempts for attackers. In cases of cache misses, the attacker employs an iterative request modification and resubmission strategy. To maintain attack accuracy and prevent false positives, avoiding interactions with previously cached content from the attacker's requests is crucial. Therefore, each constructed input query must incorporate unique prefixes, ensuring distinctive cache signatures across iterations. Notably, the attackers focus on extracting private inputs from users sharing the same cache and do not target the victim users' identification information.

\noindent\textbf{Attacker’s Capability. }We consider a realistic attacker who accesses the LLM service as an ordinary user through the public interfaces provided by the cloud service. The attacker does not possess special privileges and is unaware of the training data, model parameters, or specific hardware configurations. The attacker's capabilities are confined to sending requests and measuring the response times, as permitted by the API.

\subsection{Challenges}

The controlled experiments enable us to analyze cache-related timing characteristics in an environment free from remote API and establish a foundation for understanding exploitable timing patterns in real-world scenarios. While adversaries can theoretically craft input prompts to align with cached content from other users, several significant challenges emerge in real-world environments.

\noindent\textbf{Expansive Search Space. }Modern language models may incorporate vocabulary sizes exceeding 100,000 tokens~\cite{tao2024scaling}, with context windows extending to two million tokens~\cite{ding2024longrope}. This exponential growth in search space with increasing input length renders accurate input reconstruction computationally intractable, presenting a fundamental challenge to effective input construction. Moreover, the diversity of natural language expressions and the potential variations in user prompts further expand this search space, making exhaustive exploration impractical even for relatively short prompts.

\noindent\textbf{Interference from Time Noise. }Response time measurements are subject to various environmental factors, including network latency and memory scheduling delays. These external influences introduce noise that obscures the true response time, potentially compromising the accuracy of cache status judgment. Additional factors such as load balancing, resource contention, and system load variations can further complicate timing measurements in distributed systems. The challenge becomes particularly acute in cloud-based deployments where multiple layers of virtualization and shared resources introduce additional timing uncertainties.

\noindent\textbf{Real-world Constraints. } Several constraints complicate our attack. Firstly, rate limits are commonly imposed to prevent API abuse and ensure fair service for all users, restricting the number of attempts we can make when interacting with the API. Secondly, limited GPU memory and the costs associated with API usage constrain the total number of tokens an attacker can try, preventing continuous attempts that might evict the target request from memory. Finally, as noted by OpenAI, cached content typically remains active during inactivity periods lasting from 5 to 10 minutes and can persist for up to one hour during off-peak periods~\cite{OpenAIPrefixcaching}. Consequently, the duration of the end-to-end attack is limited by the cache's Time To Live (TTL). Additionally, dynamic pricing models and usage quotas implemented by service providers further restrict the feasibility of large-scale attack attempts, while sophisticated rate-limiting algorithms may detect and block suspicious patterns of API usage.
\section{Attack 1:  Prompt Engineering}

This section introduces our input theft attacks in applications assisted by prompt engineering optimized with the prefix caching mechanism. We will demonstrate the effectiveness of our attacks through experiments, highlighting the privacy information leakage risks associated with LLM-based applications.

\subsection{Introduction to Attack Scenario}

Prompt engineering may influence the generative performance of LLMs ~\cite{sclar2023quantifying,marvin2023prompt}. The parameterized prompt is a format of prompt template that enables dynamic value substitution based on user input~\cite{gim2024prompt}, which has been advocated in applications such as GPTforWork~\cite{gptforworkOpenAIPrompt} and various prompt generation utilities~\cite{generatestory,webutility}. In our attacks, the systems integrate user-provided information into predefined positions within hidden prompts, facilitating personalized response generation. 

\begin{figure}[!htbp]
    \centering    
    \includegraphics[width=0.95\linewidth]{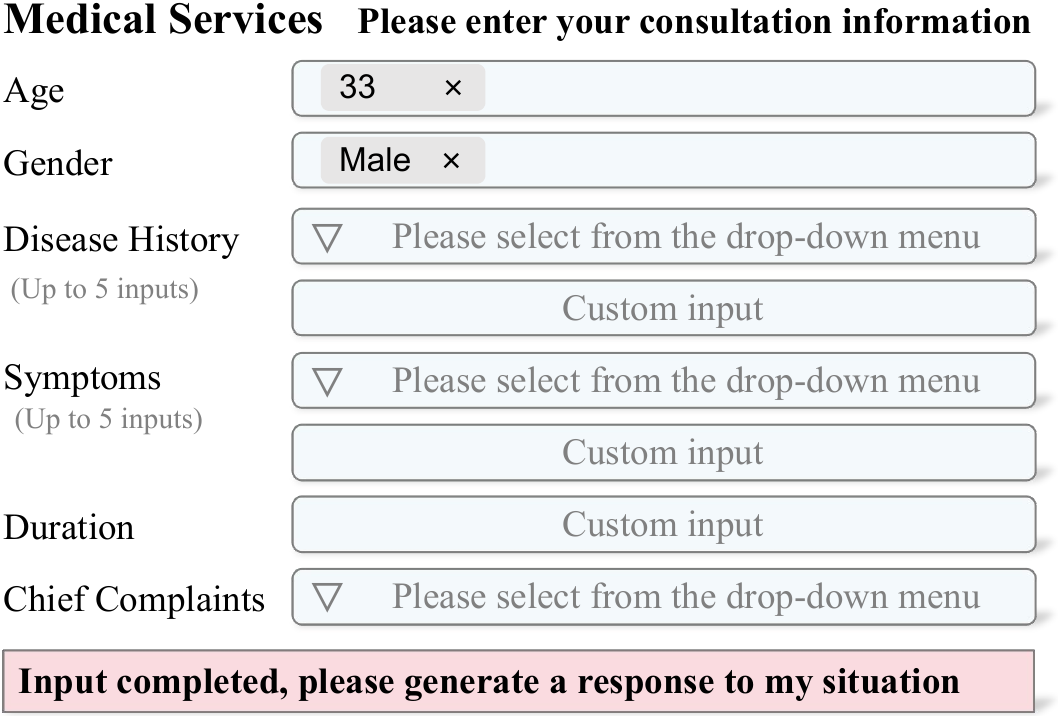}
    \caption{Medical consultation interface. Collecting user information through six fields, featuring structured selections and flexible custom inputs, supporting predefined options and free-form text entry for friendly user interaction.}
    \label{fig:  medical}
\end{figure}

User interaction involves sensitive data submissions, such as medical records, personal preferences, financial information, and proprietary business data. We focus on a medical consultation system inspired by the Zuoshou doctor platform~\cite{zuoshouyisheng}, where users provide sensitive input across six fields illustrated as Figure~\ref{fig:  medical}. While \textit{Age} and \textit{Gender} fields have numerical constraints, the \textit{Disease History} and \textit{Symptoms} fields have unlimited amounts with user-customizable items. The system maintains consistent item ordering when populating the prompt template based on research indicating the impact of input sequencing on generation quality~\cite{chu2024better}. \textit{Duration} field accepts free-form input. \textit{Chief Complaints} offers ten predetermined options based on the model's capabilities, covering aspects such as treatment approaches, medication management, dietary considerations, etc. For standardization purposes, both \textit{Disease History} and \textit{Symptoms} are restricted to 100 characters, with placeholder text automatically filling any unused capacity.

\subsection{Attack Methodology}

\begin{figure}[!htbp]
    \centering    
    \includegraphics[width=0.95\linewidth]{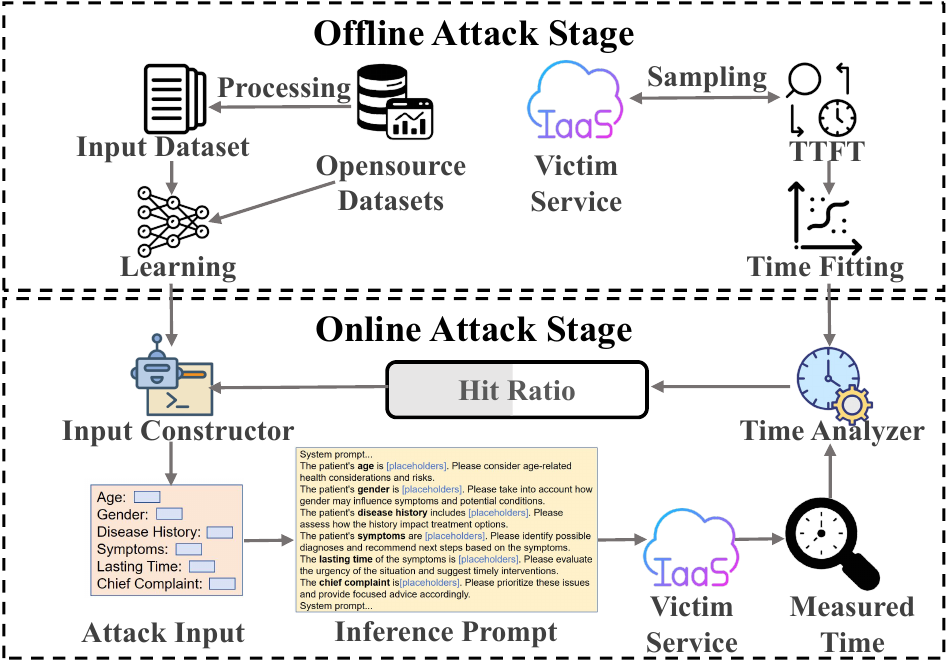}
    \caption{Attack overview. The input constructor leverages field correlation learning to optimize online input generation, while the timing analyzer establishes temporal patterns during the offline phase to facilitate online timing analysis. }
    \label{fig:  attack1}
\end{figure}

Our attack framework comprises two primary components:  the \textit{time analyzer} and the \textit{input constructor}, as illustrated in Figure~\ref{fig:  attack1}. During the offline attack phase, the input constructor learns the relationships between input fields from open-source datasets, and the timing analyzer obtains the association between response time and hit ratios by the appropriate query. In the online attack phase, the constructor generates the inputs embedded into the system prompt for LLM inference. The response time is then measured and fed into the time analyzer to determine whether the current field is hit.

The inference backend, exemplified by vLLM, employs block-level memory management for the KV Cache. Specifically, vLLM allocates and manages KV Cache memory in fixed-size blocks, enabling efficient memory utilization and cache management. Different LLM service providers implement varying block sizes, OpenAI utilizes 128-token blocks~\cite{OpenAIPrefixcaching}, while DeepSeek employs 64-token blocks~\cite{DeepSeekapi}. The standardized block-based cache allows us to verify cache hits on a block granularity, making it possible to conduct inputs field by field sequentially. Once we successfully hit one input field, the next field can be constructed given the context, significantly reducing the search space and improving attack efficiency.

\begin{figure}[!htbp]
    \centering    
    \includegraphics[width=0.95\linewidth]{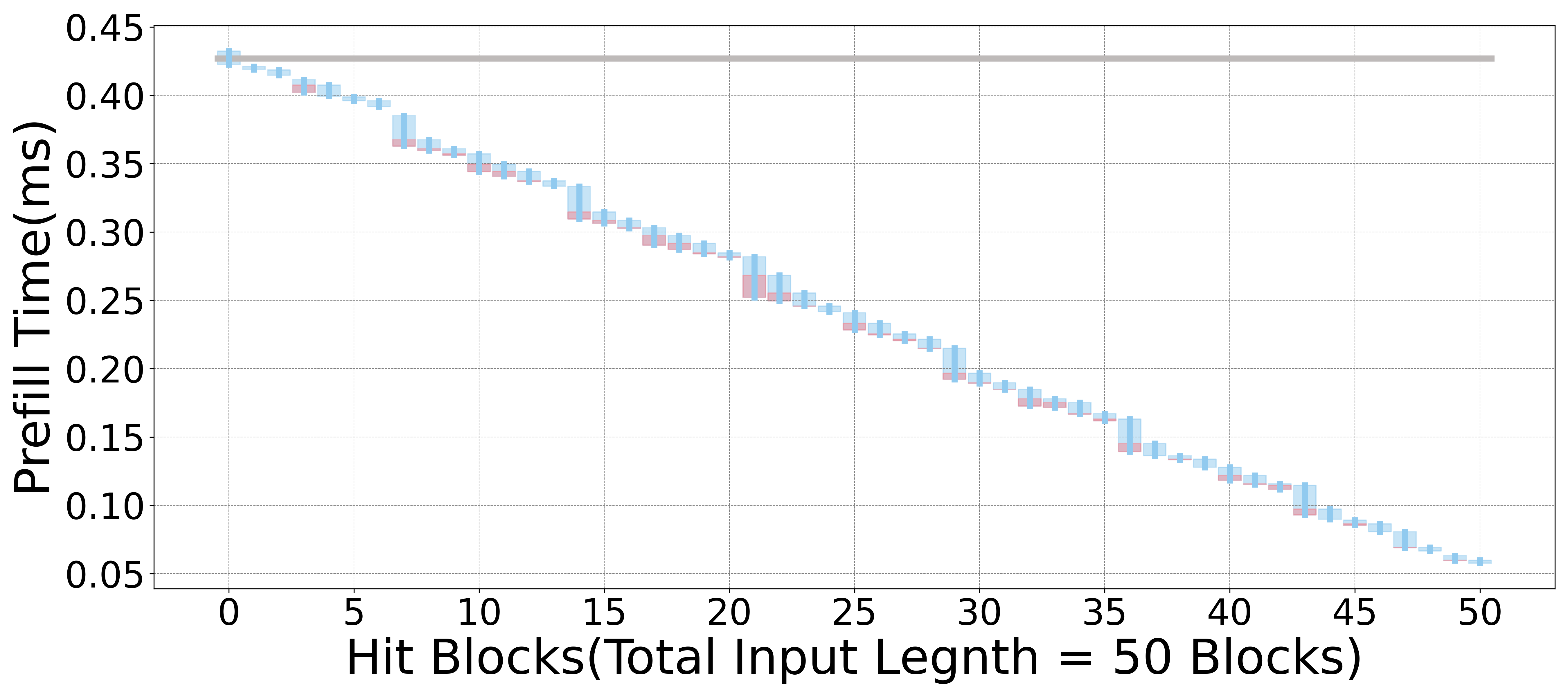}
    \caption{Relationship between TTFT and hit blocks. A total of 30 experiments were conducted with a prompt length of 800 tokens, where each block consists of 16 tokens. The gray line indicates the prefill time no block hits and the blue section represents the time range for different hit blocks obtained from multiple experiments. The red section indicates the overlapping time intervals for different hit block counts, which are prone to misjudgment.}
    \label{fig:  800}
\end{figure}

\noindent\textbf{Time Analyzer.} Through 30 experimental trials, we analyzed temporal distribution patterns, revealing distinct response time variations corresponding to different cache block hit counts. As illustrated in Figure~\ref{fig:  800}, the blue regions represent temporal distributions for current hit blocks, with distinguishable time intervals enabling hit ratio estimation for field-level construction. The red regions highlight temporal overlaps between different hit ratios, presenting a significant challenge as similar response times may correspond to multiple distinct hit ratios. These overlapping patterns predominantly occur between adjacent hit block counts and are exacerbated by temporal noise arising from concurrent request processing and hardware scheduling variations. This temporal ambiguity constitutes a fundamental challenge to our attack methodology, necessitating the development of robust prediction strategies.

The prefill computation time exhibits a relationship concerning the prompt length and hit ratio, expressed as time is proportional to (n-k) × n, where n denotes the total token count and k represents hit token counts. Assuming minimal system noise, this relationship enables precise prediction of hit ratios based on observed response times. During the offline phase, we conduct sampling of the target service to get this functional relationship, employing post-processing techniques to filter anomalous measurements. The noise is inherent and unavoidable due to system variations and concurrent request interference. Instead of making deterministic predictions, we maintain multiple candidate hit ratios weighted by their likelihood given the observed time. This statistical framework proves effective in practice since adjacent input fields are separated by more than one block length, and field content contains multiple blocks. These structural input characteristics help isolate potential misclassifications, preventing error propagation from blocks to fields and ensuring that occasional block-level prediction errors do not significantly impact the overall attack success rate.

\begin{figure}[!htbp]
    \centering    
    \includegraphics[width=0.95\linewidth]{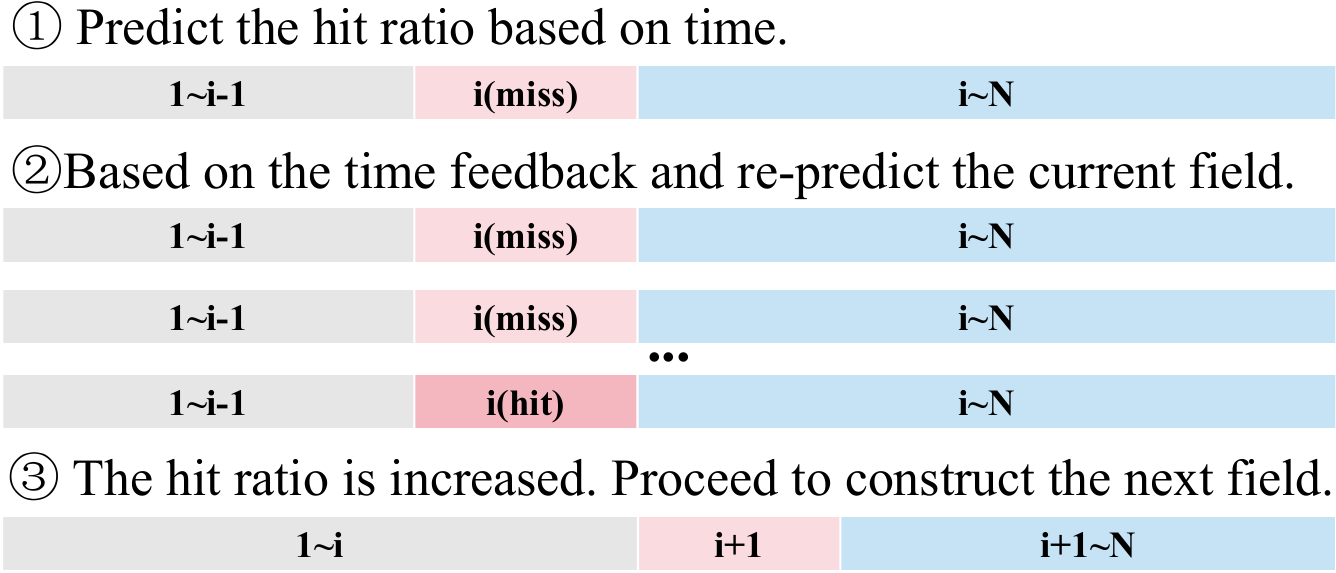}
    \caption{Partial prediction mechanism with block ratio estimation. The construction of the next field can only proceed after hitting the current field.}
    \label{fig:  partial_predict}
\end{figure}

\noindent\textbf{Input Constructor.} The block-level management mechanism enable input constructor to construct input field-by-field, as illustrated in Figure~\ref{fig:  partial_predict}. When constructing the $i-th$ field, the preceding fields remain unchanged, and the following fields are randomly filled. Subsequent fields are constructed if the preceding fields hit. If the current field fails to hit, the input constructor continues to generate different alternative content to avoid erroneous successful judgments caused by hitting previous attack inputs. Upon a successful hit of the current field, the process advances to constructing the next field.

The attacker lacks fine-grained information corresponding to user inputs, and exhaustively searching the vocabulary space for each position would result in a prohibitively large search space. To reduce this complexity, we leverage open-source datasets to learn potential field contents and their probabilities. We analyze these datasets to construct a targeted vocabulary subset for each field based on its semantic context and typical usage patterns. This context-aware approach substantially narrows the search space by considering only contextually relevant tokens rather than the entire vocabulary. Furthermore, we can reduce the search space for subsequent fields by combining hit inputs with the learned field correlations, making the attack computationally feasible. For example, a specific age or gender is often strongly correlated with certain diseases, and known disease information can aid in constructing subsequent symptoms. 

\subsection{Experimental Setup}

\noindent\textbf{Experimental Setup. }
We conduct experiments on vLLM 0.6.2 as the LLM inference framework, which implements block-based memory management and prefix caching. We simulate multi-user scenarios by invoking the OpenAI-compatible local API through various processes. The attacker continuously attempts to construct input to hit cached content from other users within the same computing node. We deploy LLaMA-2 70B as the service model on an 8×A40 GPU cluster (40GB memory per GPU), representing a typical production deployment. The model is configured with deterministic sampling parameters:  $temperature$=0, $top_k$=1, and $top_p$=1.

\noindent\textbf{Datasets. }
We learn domain-specific vocabulary and inter-field correlations from open-source Chatdoctor~\cite{li2023chatdoctor} datasets, which comprise 110K real patient-doctor conversations. We extract age-containing dialogue samples from the patient's input and employ ChatGPT-4o to extract and structure information across six distinct fields to get our Formatted Input Field datasets containing 16,276 samples. To facilitate large language model fine-tuning, we processed three customized fields, where the input consists of successfully predicted fields and the output corresponds to the target field for prediction. This processing resulted in three datasets of equal size, each designed to train the model on specific field prediction tasks while leveraging the information from previously identified fields. Our complexity analysis reveals that the search space of the Formatted Input Field dataset exceeds $2\times10^{42}$, highlighting the computational challenges involved. 

To ensure comprehensive evaluation, we implement post-processing validation mechanisms for maintaining structural consistency while preserving semantic integrity, specifically addressing format inconsistencies and field omissions in large language model outputs. The evaluation methodology incorporates 200 randomly selected samples from the Formatted Input Field dataset to simulate real-world user inputs, with the remaining data allocated for constructor training. Additionally, to enhance evaluation fairness and assess generalization capabilities, we supplemented our test set with 200 ChatGPT-4 generated samples, simulating diverse user expressions and ensuring unbiased performance assessment.

\noindent\textbf{Evaluation Metrics. }Our evaluation framework encompasses both idealized scenarios and practical constraints. The system operates within a memory allocation of 128GB for model parameters, with 80\% of the remaining memory (153GB) dedicated to KV-cache storage, accommodating 250K tokens' worth of KV vectors. This configuration establishes our upper bound for total input token capacity. We implement a pragmatic 5-minute attack duration limit, which aligns with typical prefix cache lifetimes in production environments~\cite{OpenAIPrefixcaching}. For rate limiting, we adopt the Tier2 user constraints from OpenAI's specification of 5,000 requests per minute (RPM), while acknowledging that rate limits vary substantially across different API providers, with some offering unrestricted access. We evaluate stealing success rates for disease ($ASR_{disease})$ and symptom ($ASR_{symptoms}$) prediction, overall input stealing success rates ($ASR_{all}$), number of attempts($Attempts$), required memory utilization($Tokens$) and time consumption($Time$). 

The experimental evaluation encompasses two distinct scenarios:  an unconstrained setting ($Ideal$) representing theoretical maximum performance, and a constrained setting ($All$) that incorporates all three practical limitations (memory, time, and rate constraints). This dual-scenario evaluation framework enables us to quantify both the theoretical upper bounds of system performance and its practical efficacy under real-world operational constraints, where the performance delta between scenarios provides valuable insights into constraint impacts and highlights potential optimization opportunities.  

\subsection{Evaluation Results}
\noindent\textbf{Timing Analyzer Evaluation. }
We evaluate our timing analyzer on prompts of 800 and 1,600 tokens as shown in Table~\ref{tab: time_800} and Table~\ref{tab: time_1600}. For each prompt length, we conduct experiments with varying sampling quantities (10, 300, 600, and 900 samples) and evaluate three machine learning methods:  Gradient Boosting, Random Forest, and XGBoost. The models are tested on a dataset of 1,000 samples, measuring two key metrics:  $SR_{Block}$ (success rate in predicting the number of cache-hit blocks) and $SR_{Field}$ (success rate in predicting hits for four fields, assuming field starting positions are separated by at least 16 tokens, which aligns with realistic usage patterns). This comprehensive evaluation framework allows us to assess the impact of sampling quantity and the effectiveness of different prediction methods.

The results demonstrate that our timing analyzer can achieve high accuracy with moderate sampling overhead, making it efficient for real-world attacks. As shown in Table~\ref{tab: time_800}, for 800-token inputs, our analyzer achieves optimal performance with 600 sampling points, demonstrating an 86.34\% success rate in predicting cache-hit block counts and a near-perfect 97.25\% accuracy in field-level hit detection. Notably, increasing sampling beyond this point yields diminishing returns, indicating that moderate sampling quantities are sufficient for effective timing analysis. Similar patterns emerge for 1600-token inputs as shown in Table~\ref{tab: time_1600}, where 600 samples achieve optimal performance with 87.13\% accuracy in block hit prediction and 100\% accuracy in field-level hit detection. 

\begin{table}[h]
    \centering
    \caption{Performance of the Time Analyzer With Input Token Number is 800}
    \label{tab: time_800}
\begin{tabular}{|c|l|l|l|}
        \hline
{Query Numbers} & {Prediction Methods} & $SR_{Block}$ & $SR_{Field}$ \\
        \hline

\multirow{3}{*}{100}               & Gradient Boosting                     & 68.43\%            & 95.42\%            \\
                                  & Random Forest                          & 64.64\%            & 94.58\%            \\
                                  & XGBoost                                & 56.27\%            & 95.88\%   \\
        \hline
\multirow{2}{*}{300}              & Gradient Boosting                      & 81.83\%            & 96.41\%   \\
                                  & Random Forest                          & 78.76\%            & 95.69\%            \\
                                  & XGBoost                                & 72.55\%            & 96.01\%   \\
        \hline
\multirow{3}{*}{600}              & Gradient Boosting                      & \textbf{86.34\%}   & 97.25\%   \\
                                  & Random Forest                          & 79.28\%            & 96.21\%            \\
                                  & XGBoost                                & 81.63\%            & 96.67\%            \\
        \hline
\multirow{3}{*}{900}              & Gradient Boosting                      & 80.13\%            & \textbf{97.45\% }  \\
                                  & Random Forest                          & 81.18\%            & 96.86\%            \\
                                  & XGBoost                                & 82.35\%            & 96.80\%            \\
        \hline
\end{tabular}
\end{table}

\begin{table}[h]
    \centering
    \caption{Performance of the Time Analyzer With Input Token Number is 1600}
    \label{tab: time_1600}
\begin{tabular}{|c|l|l|l|}
        \hline
{Query Numbers} & {Prediction Methods} & $SR_{Block}$ & $SR_{Field}$ \\
        \hline
\multirow{3}{*}{100}              & Gradient Boosting                      & 48.51\%            & 93.07\%            \\
                                  & Random Forest                          & 43.47\%            & 91.29\%            \\
                                  & XGBoost                                & 36.04\%            & 95.05\%   \\
        \hline
\multirow{3}{*}{300}              & Gradient Boosting                      & 75.84\%            & 100\%            \\
                                  & Random Forest                          & 74.95\%            & 100\%            \\
                                  & XGBoost                                & 64.06\%            & 100\%            \\
        \hline
\multirow{3}{*}{600}              & Gradient Boosting                      & \textbf{87.13\%}   & 99.70\%            \\
                                  & Random Forest                          & 84.65\%            & \textbf{100\% }           \\
                                  & XGBoost                                & 79.31\%            & \textbf{100\%}            \\
        \hline
\multirow{3}{*}{900}              & Gradient Boosting                      & 86.73\%            & 100\%            \\
                                  & Random Forest                          & 84.54\%            & 100\%            \\
                                  & XGBoost                                & 82.97\%            & \textbf{100\%  }          \\
        \hline
\end{tabular}
\end{table}

\noindent\textbf{End-to-End Attack Evaluation.} We integrate our input constructor and time analyzer for end-to-end attack assessment in a shared prefix KV cache, where regular users and attackers coexist on the same node. Attackers can determine cache hits and infer other users' inputs by analyzing TTFT patterns of constructed requests. The attack proceeds by conducting exhaustive searches for basic fields like age and gender, then leveraging these successfully matched fields to sequentially predict subsequent entries, with custom content fields presenting the highest complexity.

We implement four approaches to implement the input constructor. (1) The baseline approach employs random construction based on knowledge learned from  Formatted Input Field datasets. (2) Additionally, we developed a machine learning method using Gaussian Naive Bayes, comprising two random constructors for Age and Gender and four distinct models trained on Formatted Input Field for subsequent field prediction. (3) The framework also incorporates probabilistic learning implemented at fine-grained vocabulary levels, focusing on vocabulary probabilities and combination relationships. (4) We utilize fine-tuned LLaMA-3 8B models trained on three Formatted Input Field datasets, specifically targeting high-complexity fields (Disease History, Symptoms, Duration). Each model specializes in predicting its designated field based on previously matched inputs. To ensure diverse outputs match user input patterns, we configure the models with high sampling parameters (temperature=0.99, top\_p=0.99, max\_tokens=100) and generate 30 candidates per inference. Given the inherent output diversity and potential duplications, we perform 50 inference iterations for each test sample, remove duplicates, and employ a ranking mechanism to prioritize the predictions. These ranked outputs are then sequentially submitted to the victim service for cache hit attempts.

The comprehensive results are presented in Table~\ref{tab: e2e_exp}. The experimental results demonstrate the performance variations across different scenarios and methods. Under ideal conditions, the Probability-based Vocabulary approach demonstrates superior performance, achieving the highest attack success rates across disease prediction (67.5\%), symptom prediction (53.75\%), and overall field prediction (49\%). However, due to its word-level search mechanism and relatively poor grasp of coarse-grained correlations, this method incurs substantial computational overhead. In contrast, the GaussianNB method strikes an optimal balance between resource utilization and performance, showing approximately twofold efficiency improvement compared to the baseline approach. When subjected to real-world constraints, while the Probability-based Vocabulary method maintains its leadership in disease prediction (62\%), its advantages in symptom and overall prediction become less pronounced. The GaussianNB method exhibits remarkable stability, achieving the highest overall prediction accuracy (12.50\%).

The Finetuned LLM approach, hampered by output variability and uncertainty, requires extensive attack attempts, resulting in excessive memory and time consumption, yielding performance dropping from 54.00\% in ideal conditions to complete failure under comprehensive constraints.  These findings suggest that traditional machine learning methods offer a more reliable and efficient solution for precise matching tasks under practical constraints than large language models, particularly when considering the trade-off between accuracy, resource utilization, and operational stability. The Finetuned LLM shows promising potential in unrestricted environments, its substantial resource requirements and performance instability under constraints make it less suitable for real-world attacks.

\begin{table*}[htbp]
    \centering
    \caption{End-to-End Attack Performance Comparison Across Different Methods Under Various Constraints. We evaluate success rates for disease prediction ($ASR_{disease}$), symptom prediction ($ASR_{symptoms}$), and overall field prediction ($ASR_{all}$), along with required attempt counts($Attempts$), memory usage ($Tokens$), and time consumption ($Time$). The evaluation is conducted under two scenarios:  $Ideal$ represents the theoretical upper bound without any practical limitations, $All$ reflects real-world performance under memory constraints (250K tokens), rate limits (5,000 RPM), and time restrictions (5-minute duration).}             
    \label{tab: e2e_exp}
\begin{tabular}{|l|r|r|r|r|r|r|}
        
        \hline
        \multicolumn{7}{|c|}{Ideal}\\
        \hline
        {Methods} & {$ASR_{disease}$} & {$ASR_{symptoms}$} & {$ASR_{all}$} & {Attempts} & {Tokens} & {Time} \\
        \hline
        Baseline         & 60.00\%  & 23.50\% & 22.00\%   & $6868 \pm 6855$      & $283799 \pm 273969$     & $1751 \pm 1747$\\
        \hline
        
        GaussianNB       & 59.00\%  & 27.50\% & 26.50\%  & $2904 \pm 2830$      & $108650 \pm 94450$      & $730 \pm 622$\\
        \hline

        Prob\_vocabulary & 67.50\%   & 53.75\% & 49.00\%  & $502720 \pm 2426519$ & $11507996 \pm 56966635$ & $77918 \pm 376100$ \\
        \hline

        Finetuned LLM    & 54.00\%  & 18.00\% & 10.00\%  & $659920 \pm 6817828$ & $6599206 \pm 6817828$   & $101500 \pm 101001$ \\
        \hline
        
        \multicolumn{7}{|c|}{All}\\
        \hline
        {Methods} & {$ASR_{disease}$} & {$ASR_{symptoms}$} & {$ASR_{all}$} & {Attempts} & {Tokens} & {Time} \\
        \hline
        Baseline           & 45.50\% & 9.00\%  & 8.00\%  & $1350\pm 1173$  & $55926 \pm 48139$ & $ 232 \pm 166 $\\
        \hline

        GaussianNB         & 54.00\% & 13.50\% & 12.50\% & $1086\pm 898$   & $40207 \pm 33421$ & $ 276\pm 228 $\\
        \hline

        Prob\_vocabulary   & 62.00\% & 12.50\% & 9.50\%  & $2247\pm 983$   & $57526 \pm 26768$ & $ 348\pm152 $\\      
        \hline
        
        Finetuned LLM      & 00.00\% & 00.00\% & 00.00\% & $0\pm 0$        & $0 \pm 0$         & $ 0\pm 0$\\
        \hline
\end{tabular}
\end{table*}

\section{Attack 2:  Retrieval Augmented Generation}

This section introduces our input theft attacks, which target semantic caching in applications assisted by retrieval-augmented generation. We will demonstrate the effectiveness of our attacks through experiments, highlighting the privacy information leakage risks associated with LLM-based applications.

\subsection{Introduction to Attack Scenario}
Semantic caching represents a crucial optimization for LLM and RAG applications, offering developers significant practical advantages. Research indicates that up to 31\% of LLM calls are redundant, which can be effectively eliminated through semantic caching ~\cite{gill2024privacy}. This optimization is particularly vital in Retrieval-Augmented Generation (RAG) systems, where rapid and accurate response generation is critical for performance. Intelligently caching semantically similar queries substantially reduces database retrieval and LLM inference while dramatically improving response times. For application developers, this translates into more cost-effective scaling, improved system reliability, and enhanced user experience. The growing adoption of semantic caching is evident in LLM-based tools like GPT for Work~\cite{gptforwork} plugin, attracting 6.88 million users. As LLM applications continue to grow in popularity, semantic caching emerges as an essential architectural pattern that effectively bridges the gap between powerful LLM capabilities and real-world deployment constraints.

We identify a novel privacy vulnerability in RAG-assisted LLM systems with semantic caching. In contrast to exact matching approaches, our study examines semantic caching systems that leverage similarity matching mechanisms, where responses are cached and retrieved based on semantic proximity rather than exact matches. Through systematic probing with carefully crafted queries, attackers can exploit the timing side-channel to infer the presence of specific topics or questions in the cache, potentially revealing sensitive information about other users' interactions with the system.

We demonstrate our attack scenario in the context of online legal consultation services, where RAG-enhanced LLM systems are commonly deployed to supplement responses with up-to-date regulations and relevant case law. These services typically implement semantic caching to optimize performance and reduce computational overhead when handling similar inquiries from multiple users. By exploiting the shared cache architecture, our attack framework targets legal consultation queries spanning various domains, enabling attackers to infer the semantic content and topical nature of other users' legal inquiries. This represents a significant privacy concern as it exposes sensitive information about users' legal circumstances and consultation topics.

\subsection{Methodologies}
Our attack methodology comprises two key components:  a input constructor and a time analyzer. Due to the reduction in retrieval and LLM inference processes when semantic responses are cached, there is a significant temporal disparity between cache hits and misses, as illustrated in Figure~\ref{fig:  gptcache_exp}. Through temporal feature extraction from sampled data, our timing analyzer achieves 100\% accuracy in distinguishing between cache hits and misses. However, the input constructor faces substantial challenges in crafting inputs within an infinite search space, as it must generate semantically relevant queries that effectively probe the cache contents.

We propose an intelligent constructor that systematically explores an extensive, unknown input space to hit the cached user input. Our approach begins with semantic space partitioning through hierarchical clustering of the training dataset, establishing a foundational understanding of the input space structure. To efficiently navigate this partitioned space, we implement a weighted binary tree whose weights are derived from cluster cardinality, enabling depth-first exploration of semantic clusters. The search strategy carefully balances exploitation and exploration. While focusing on promising regions near cluster centroids to leverage learned patterns, the constructor simultaneously conducts peripheral exploration around clusters to maintain search breadth. This dual-focus approach is governed by parameters that dynamically adjust the exploration-exploitation trade-off based on search progress.

In each iteration, the constructor generates multiple candidate inputs, subjecting them to a sophisticated ranking process. This ranking mechanism integrates multiple factors:  historical attempt patterns, similarity relationships with previous attempts, and the candidate's representativeness within the current pool. To maintain attack efficiency and prevent redundant attempts, the constructor enforces diversity constraints through similarity thresholds, ensuring each new attempt is sufficiently distinct from previous ones while remaining representative of the current search region.

The effectiveness of our constructor lies in its ability to systematically explore high-dimensional semantic spaces while adapting its search patterns based on accumulated knowledge. Through this balanced approach of focused exploitation and strategic exploration, the constructor efficiently navigates the challenge of matching user inputs in an extensive unknown space while maintaining diversity in its attack attempts.

\subsection{Experimental Setup and Evaluation}
\noindent\textbf{Experimental Setup. }
Our experimental framework utilized GPTCache's semantic caching infrastructure. We configured the semantic similarity thresholds as 0.9 using its default similarity evaluation method \textit{SearchDistanceEvaluation}~\cite{gptcacheBetterConfigure}. We tend to choose a deliberately conservative threshold setting to ensure high-fidelity semantic matching and to impose stringent quality requirements on our constructor's output. This high threshold helped eliminate false positives in cache hits, enabling more precise measurement of performance differences. The RAG database was constructed by uploading legal corpus to the OpenAI organization platform, where we measured and analyzed retrieval latencies under various query inputs.

\noindent\textbf{Input Constructor Evaluation.} 
Our study employed a large-scale legal consultation corpus CrimeKgAssitant~\cite{CrimeKgAssitant} consisting of 200,000 question-answer pairs, categorized into 13 distinct legal domains. The dataset's categorical distribution, training data sizes, and input length statistics are detailed in Table~\ref{tab: law_data}. We extracted the user query components from the dataset for input constructor training. To simulate diverse real-world user inputs, we employ ChatGPT-4 to generate 30 test queries and split 30 test data from the training dataset, forming a test set that reflects varied user expression patterns.

Empirical results demonstrate that our prompt constructor can effectively capture semantic spaces from training data. Through iterative probing of user input requests, it achieves semantic extraction success rates ranging from 43\% to 100\%, as presented in Table~\ref{tab: law_data}. This effectiveness in generalizing across disparate datasets highlights the robust learning capabilities in input reconstruction.

\begin{table}[htbp]
    \centering
    \caption{Systematic evaluation of attack success rates (ASR). Constructor learned from CrimeKgAssitant datasets with 13 legal domains and tested with GPT-4 generated data}
    \label{tab: law_data}
    \begin{tabular}{|l|c|c|c|c|}
        \hline
        \textbf{Category} & \textbf{Count} & \textbf{Len. Avg. $\pm$ Std.} & \textbf{ASR} \\

        \hline
        Marriage and Family & 39281 & 38.11 $\pm$ 344.29 & 93.33\% \\
        \hline
        Labor Disputes & 35011 & 38.79 $\pm$ 329.54 & 90.00\%  \\
        \hline
        Traffic Accidents & 22646 & 39.45 $\pm$ 317.98 & 100.00\% \\
        \hline
        Debt Disputes & 21925 & 38.47 $\pm$ 364.97 & 90.00\%  \\
        \hline
        Criminal Defense & 18314 & 35.96 $\pm$ 364.64 & 93.33\% \\
        \hline
        Contract Disputes & 13765 & 39.73 $\pm$ 340.30 & 96.67\% \\
        \hline
        Property Disputes & 12071 & 36.65 $\pm$ 337.62 & 100.00\%  \\
        \hline
        Infringement & 10594 & 39.03 $\pm$ 426.91 & 43.33\%  \\
        \hline
        Company Law & 10011 & 38.47 $\pm$ 340.90 & 70.00\%  \\
        \hline
        Medical Disputes & 7285 & 39.07 $\pm$ 345.55 & 73.33\% \\
        \hline
        Demolition and Resettlement & 7022 & 40.91 $\pm$ 320.75 & 100.00\% \\
        \hline
        Administrative Litigation & 2776 & 36.14 $\pm$ 302.39 & 93.10\% \\
        \hline
        Construction Projects & 1610 & 42.11 $\pm$ 262.82 & 63.33\% \\
        \hline
    \end{tabular}
\end{table}

\section{Discussion and Defenses}
\subsection{Discussion}
\noindent\textbf{Trade-offs in Cloud-Deployed LLMs.}
While local LLM deployment provides inherent privacy benefits, cloud deployment remains inevitable due to the substantial computational requirements of modern inference systems that exceed edge device capabilities. This fundamental tension between privacy and performance has led to various privacy-preserving solutions, including homomorphic encryption, TEEs, and data masking for cloud-based inference. However, our timing side-channel attack demonstrates that these cryptographic approaches fail to address privacy leakage arising by system-level optimizations like prefix cache-sharing.

Our research exposes a critical vulnerability where performance optimizations in cloud-deployed LLMs, while essential for scalability, create exploitable privacy side channels. This finding reveals an important trade-off:  shared caching mechanisms that enhance system efficiency simultaneously introduce subtle yet significant privacy risks. The implications extend beyond LLM systems, contributing to the broader discourse on balancing security and performance in cloud computing architectures and emphasizing the need for comprehensive privacy-preserving designs that consider cryptographic protections and system-level vulnerabilities.

\noindent{\textbf{Limitations of the Attack.}}
One significant limitation of our attack is the lack of fine-grained information for accurately reconstructing the input. While our insights effectively capture secret information in the data, it does not provide the detailed context required for precise recovery. To address this, integrating techniques such as embedding inversion and token-length side channels could enhance the attack's effectiveness by offering additional granular information. Our methods could serve as validation mechanisms, improving the reliability of the input recovery process.

Another critical limitation is that our attack cannot link the inferred inputs to specific users or input objects, preventing meaningful use of the stolen data in real-world scenarios. This lack of user attribution significantly limits the attack's practical implications.

\subsection{Potential Defenses}
Our attack involves manipulating prompt sequences to probe the KV Cache, allowing adversaries to determine cache hits or misses based on the time taken to generate responses. This section delineates strategic defense mechanisms to mitigate such timing attacks.

\noindent\textbf{User-Level Cache Isolation. }Implementing distinct cache namespaces prevents cross-user cache sharing, effectively containing cache states within individual sessions. While this approach strengthens security, it trades off system efficiency. Major providers like  OpenAI's API~\cite{OpenAIPrefixcaching} and DeepSeek's API~\cite{isolated_cache} implement isolation for prefix caching to protect user privacy and security.

\noindent\textbf{Rate Limiting to Mitigate Frequent Sttacks. } 
By restricting request frequency, rate limiting impedes rapid successive probing necessary for timing analysis. This defense not only prevents brute-force attempts but also maintains infrastructure stability. While OpenAI employs this approach~\cite{limit_rate}, careful calibration is needed to balance security with legitimate user access.

\noindent\textbf{Complicate Time Analysis. }Timing obfuscation approaches can effectively prevent attackers from exploiting response time variations. Similar techniques have been proven effective against model extraction attacks, where timing patterns can reveal model parameters through correlations with activation functions~\cite{batina2019csi}, network depth ~\cite{duddu2018stealing}, and computational operations~\cite{dong2019floating}. We propose two key obfuscation strategies:  response time homogenization through either constant-time execution ~\cite{maji2021leakynets} or random delay injection~\cite{breier2023desynchronization} and disabling streaming responses to eliminate measurable timing patterns

Our defense strategy emphasizes the critical balance between security, performance, and user experience. While each mechanism provides distinct protection against timing-based attacks, their implementation requires careful consideration of operational requirements and threat models. A combined approach, tailored to specific deployment contexts, offers the most robust protection against our demonstrated side-channel vulnerabilities.

\section{Related Work}

\subsection{Side-channel Attacks on AI Systems}
Side-channel attacks can exploit indirect information leakage from AI systems, such as timing~\cite{duddu2018stealing, dong2019floating, gongye2020reverse}, power consumption~\cite{wei2018know,tian2021remote}, or electromagnetic emissions~\cite{batina2019csi, yu2020deepem, horvath2023barracuda}, to infer sensitive model architectures and parameters. Additionally, cache-based side channels~\cite{hong2018security, yan2020cachetelepathy,wang2022stealthy}, memory access patterns~\cite{hua2018reversecnn,hu2020deepsniffer}, and resource contention information~\cite{ naghibijouybari2018rendered, tan2021invisible, dutta2023spy} provide further avenues for gleaning model information. Debenedetti et al.~\cite{debenedetti2024privacy} explore privacy side-channel attacks in machine learning systems, highlighting how system-level components can be manipulated to leak private information more effectively than standalone models. 

However, due to the complexity of large model systems, side-channel attacks against large model systems remain relatively rare. Weiss et al.~\cite{weiss2024your} introduce a novel token-length side-channel vulnerability affecting AI assistants, demonstrating how encrypted responses from AI Chatbots can be partially reconstructed by analyzing the length of transmitted tokens over the network. In contrast, our work proposes a new timing-based side-channel attack to capture user input and analyze cache behavior over time, revealing the cache-sharing dynamics within LLM systems.

\subsection{Prompt Text Leakage Attacks}
\noindent\textbf{Adversarial Prompts.} The art of prompt engineering is crucial and often regarded as proprietary because it optimizes model performance for specific tasks~\cite{wei2022chain}, enhances instruction-following capabilities~\cite{ouyang2022training}, and aligns outputs with human values~\cite{bai2022training}, making system prompts valuable assets~\cite{warren2023these}. Perez et al.~\cite{perez2022ignore} first explore prompt leakage in LLMs by injecting crafted prompts that misalign the model's intended goals, allowing attackers to extract sensitive information embedded in the prompts. Recent studies~\cite{zhang2024effective, yu2023assessing, liang2024my} primarily focus on developing more effective adversarial prompts to expose the system prompts of LLMs.

\noindent\textbf{Prompts Inversion.} Another type of attack involves inferring input prompts from generated content. Input text data is often represented as embedding vectors in natural language processing, especially with the proliferation of LLMs. Previous work~\cite{song2020information} has demonstrated the possibility of obtaining sensitive information of original inputs by inverting these embedding vectors. Since then, numerous studies have focused on embedding inversion attacks on language models, Morris et al.~\cite{morris2023text} recover significant personal information from clinical datasets, Li et al.~\cite{li2023sentence} generate coherent and similar sentences to the original input, and Chen et al.~\cite{chen2024text} investigate multilingual inversion attacks. 

In addition to embedding vectors, other information can also be used to reverse the input content. Morris et al.~\cite{morris2023language} suggest that the next-token probability distribution contains information about the preceding text, which can be exploited to reconstruct input prompts, highlighting the significance of residual information in model outputs. Similarly, images generated by text-to-image diffusion models~\cite{shen2024prompt, mahajan2024prompting} and texts produced by LLMs~\cite{sha2024prompt, yang2024prsa} can also be exploited to reverse-engineer text prompts. These attack techniques adversely affect the commercial benefits of the prompt marketplace and infringe on prompt engineers' property rights.

\subsection{Memorization and Privacy Risks in LLMs}
Memorization of language models can remember parts of the training data, which encompasses web crawls of personal pages, social media messages, and internal email databases, raising concerns about data copyright and privacy breaches. Pre-trained and fine-tuned language models~\cite{lehman2021does, mireshghallah2022empirical, huang2022large,mireshghallah2022memorization} also subject to these issues. Carlini et al.~\cite{carlini2022quantifying} demonstrate that larger models tend to memorize more information, highlighting the necessity of mitigating memorization as the model scale keeps growing. 

If the training data is leaked, it can result in the unauthorized exposure of confidential information, raising ethical concerns regarding consent and data ownership. Training data extraction attacks~\cite{carlini2021extracting, brown2022does} can recover sensitive training examples by querying the LLMs. Additionally, membership inference attacks~\cite{song2019auditing, mattern2023membership, meeus2024did,carlini2022quantifying} can determine whether specific data was included in the training dataset, potentially compromising the copyrights of the data owner. 

Memorization significantly threatens personal privacy by potentially retaining and recalling sensitive information from training data. Lukas et al.~\cite{lukas2023analyzing} explore privacy-utility trade-offs of using defenses such as PII scrubbing and Differentially Private training when fine-tuning language models. Kim et al.~\cite{kim2024propile} enable data subjects to determine whether their PII is at risk of disclosure through queries. Staab et al.~\cite{staab2023beyond} indicates that LLMs can automatically reason author attributes from unstructured text, greatly reducing the cost associated with privacy violation.




\bibliographystyle{plain}
\bibliography{ref} 

\end{document}